\documentclass[a4paper,pdflatex]{article}
\usepackage{graphicx,amsmath,amsfonts}% Include figure files
\RequirePackage[english]{babel}
\usepackage[T2A]{fontenc}
\usepackage[cp1251]{inputenc}
\usepackage{enumerate}
\title{\Large \bf Corrections to the higher moments of the relativistic ion energy-loss distribution beyond the Born approximation.\\
I. Z dependence of Mott's corrections}
\author{
O. Voskresenskaya}
\date{}

\begin{document}
\maketitle

\begin{center}
Joint Institute for Nuclear Research, Dubna, Moscow Region, 141980 Russia
\end{center}

\begin{abstract}

Using the Mott exact cross section for moderate relativistic energies, we calculated the corrections to the first-order Born higher moments of
the energy-loss distribution of charged particles
in a wide range of the particle charge numbers.

\end{abstract}

%%%%%%%%%%%%%%%%%%%%%%%%%%%%%%%%%%%%

\twocolumn{

\section{Introduction}
\vspace{5mm}

Stopping power (the average energy loss of a particle per unit path length measured
 e.g. in MeV/cm) is a necessary ingredient for many parts of basic science as well as for medical and technological applications
\cite{2005,2004,2006-1}.
The stopping power of a material
is described by the Bethe formula (the so-called ``Bethe's stopping power formula'')
\cite{Bethe1,Bethe2,Bethe3}.
 A non-relativistic version of this formula was found by
Hans Bethe in \cite{Bethe1}. Its relativistic version
 was obtained by him in \cite{Bethe2}.

The relativistic version of this formula for average ionization energy loss of charged
particles reads
\begin{eqnarray}
\label{Bethe}
-\frac{d\bar E}{dx}&=&2\zeta\left[\ln(E_m/I)-\beta^2\right],\nonumber\\
\zeta&=&0.1535(Z/\beta)^2Z^\prime/A,\nonumber\\
E_m&=& 2m c^2\beta^2(1-\beta)^{-1},
\end{eqnarray}

\smallskip

\noindent where $x$ denotes the distance traveled by a particle;
$E_m$ is the maximum transferrable energy to an electron in a collision with the particle
of velocity $\beta c$; $m$ is the electron mass;
$Z^\prime$ and $A$  refer to the atomic number and the weight of the absorber.
In the Bethe theory, the material is completely described by a single number,
the mean excitation potential $I$ of the absorber. Felix Bloch has shown \cite{Bloch} that the mean excitation
potential of the atoms is approximately given by $I$=$(10\,eV)\,Z'$.
If this approximation is introduced into formula (\ref{Bethe}) one obtains an expression
which is often called the ``Bethe--Bloch formula'' \cite{2008}.

The first-order Born approximation upon which the Bethe formula is based is inadequate to the task of accurately describing the energy-loss rate of highly charged particles (ions).
The importance of a higher-order-correction term $\Phi_\textrm{M}$ obtained by Nevill Mott \cite{Mott} to the relativistic Bethe formula \cite{Bethe1,Bethe2} that reads
\begin{eqnarray}
\label{BetheM}
-\frac{d\bar E}{dx}&=&2\zeta\left[\ln(E_m/I)-\beta^2 +\Phi_\textrm{M}/2\right],\nonumber\\
\Phi_\textrm{M} &=& \frac{1}{2}\Delta_\textrm{M}\left(d\bar E/dx\right)_\textrm{close},\nonumber\\
\Delta_\textrm{M}\!\!\left(\frac{d\bar E}{dx}\right)_{\!\!\textrm{close}}\!\!\!\!&=&n_0Z^\prime\!\!\int\!\!\left(\frac{d\sigma_\textrm{M}}{d\varepsilon}-
\frac{d\sigma_\textrm{B}}{d\varepsilon}\right)\!\varepsilon d\varepsilon\,,
\label{Bethe3}
\end{eqnarray}

\medskip

\noindent was pointed in \cite{GSI1}. In the above equations, $\Phi_\textrm{M}$ denotes the so-called ``Mott corrections'' (MCs),
$n_0$ is the number of atoms of the substance per unit volume, $\sigma_\textrm{B}$ and $\sigma_\textrm{M}$
signify the first-order Born and the Mott exact (calculated with the aid of the Mott phase shifts) cross sections, respectively.
The importance of $\sigma_\textrm{M}$ for the scattering of heavy ions off electrons with respect to the energy-loss straggling (``ELS'') of charged particles has been demonstrated in \cite{theor,GSI2}, and the well known relativistic Bohr result \cite{Bohr} is exceeded by up to a factor of 3.\\\\\\

The MCs are only available in a regime for which separation into ``distant collisions'',
in which electron-binding effects are included, and on ``close collisions'', for which the binding
energy of the electrons can be neglected  (namely, at large velocities compared to atomic velocities), is fair \cite{Will}\footnote{The idea of applying the Mott cross section to close collisions was suggested in \cite{Fow}.}.
They were first calculated exactly by numerical integration of the Mott exact cross section
in \cite{Morgan}. However, as the expression for the MCs in (\ref{Bethe3}) is extremely inconvenient for practical application,
obtaining convenient and accurate MCs expressions becomes significant.

In \cite{VSTT}, devoted to the analysis of the contribution of close collisions to
the average energy-loss of relativistic ions in matter, it has been shown that
 the exact MC expression can be represented in the form of quite rapidly converging series of quantities bilinear in the Mott partial amplitudes, which can be easily computed.
A comparison of this exact result with the approximate result of \cite{Ahlen}
was also considered there.
In this paper, we apply the exact MC expression from \cite{VSTT}
for computing the most common characteristics of the energy-loss distributions for moderate relativistic energies in a wide range of particle charge number ($5 \leq Z \leq 95$).

These characteristics (for instance, the ELS) are of interest not only because of their fundamental aspects, but also due to their impact on many applications in physics and other sciences.
So, accurate information on the ELS is important in the application of the Rutherford backscattering spectrometry, nuclear reaction analysis, and scanning transmission ion microscopy. As the ELS is one of the main factors limiting the depth resolution,
understanding the shape of energy-loss distribution is important in the application of particle identification. In view of these and other applications, the study of ELS in matter become an interesting topic in recent years.

Nevertheless, most researchers have concentrated on stopping power, and
only few investigators
%\cite{2001,2002,2003,2004-1,2006,2007,2013,2016,2016-1,2017}
have studied the ELS extensively [19--28].
Even less results devoted to the study of higher moments of the energy-loss rate.
However, it was recently shown that the higher moments of distributions of conserved quantities
measuring deviations from a Gaussian have a sensitivity to CP fluctuations \cite{2009}.
The article \cite{2010} reports the first measurement of higher moments of the distributions from
$\textrm{Au}+\textrm{Au}$ collisions to search for signatures of the CP.

The present communication is organized as follows:
We start (in Sec. 2) from a consideration of an exact expression for the Mott phase-shift formula. Using this expression, we get (in Sec. 3) a representation for the central moments ($\mu_{n,M}$) of the energy loss distribution in the form of  fast converging series. We also compare the computation results for $\mu_{n,M}$  (Sec. 3) and for the normalized central moments $\rho_{k,M}$ (Sec. 4) with the Born results. Finally (in Sec. 5), we summarize our findings.\\

\section{Evaluation of the Mott exact phase-shift formula}

\vspace{5mm}

In the higher moments of the particle energy-loss distribution that determine its shape
\cite{IDDRS}, the dominant contribution is made by close collisions. The description of the lattеr in the framework of the Born approximation
 is valid only under condition $Z\alpha/\beta\ll 1$, where
$\alpha\approx 1/137$ is the fine structure constant, and
$\beta=v/c$ is the relative ion velocity in the laboratory frame that is expressed in units of the speed of light $c$.
In calculations of the characteristics of the energy-loss distribution of heavy relativistic ions,
we should expect significant deviations from the results in the Born approximation.

When we restrict consideration by moderate relativistic energies of incident particles
such that one can neglect the effects of their electromagnetic structure,
we can use in our calculations the expression for the scattering cross section in terms of the Mott phase shifts \cite{Mott}.

As shown in \cite{VSTT}, the Mott exact phase-shift formula
can be represented as

\begin{eqnarray}
\label{sigma1}
\frac{d\sigma_\textrm{M}}{d\Omega}&=&\left(\frac{\hbar}{2p}\right)^2\frac{\left[\xi^{2}\vert
F(\vartheta)\vert^{2}+
\vert G(\vartheta)\vert^2\right]}{\sin^2(\vartheta/2)}\,,\\
F(\vartheta)&=&\sum_{l=0}^{\infty}F_lP_l(x),\; F_l=lC_l-(l+1)C_{l+1}, \nonumber\\
&&C_l=\frac{\Gamma(\rho_l-i\nu)}{\Gamma(\rho_l+1+i\nu)}e^{i\pi(1-\rho_l)}, \nonumber\\
&&\rho_l=\sqrt{l^2-(Z\alpha)^2},\;\;\nu=Z\alpha/\beta\,,\nonumber\\
G(\vartheta)&=&-\cos(\vartheta/2)\,F^\prime(\vartheta),\quad x=\cos(\vartheta)\,,\nonumber\\
&&p=mc\beta/\sqrt{1-\beta^2},\;\xi=\nu\sqrt{1-\beta^2},\nonumber
\end{eqnarray}\\
where $P_l(x)$ denotes the Legendre polynomial of order $l$ with  $x=\cos \vartheta$, $\vartheta$ is the scattering angle of the electron scattered by an ion in the center-of-mass
system, $F^\prime(\vartheta)\equiv d F(\vartheta)/d\vartheta$, and $\Gamma$ is the Euler gamma function.

Below we apply the above $d\sigma_\textrm{M}/d\Omega$ representation to calculate
the relative Mott corrections to the first-order Born central moments  and normalized central moments of the relativistic ion energy-loss distribution.

\section{
Computation of the central moments of the energy-loss distribution}

Remind now a few basic definitions. We define
the n-th moment of a real-valued continuous function $f(y)$ about a value $\bar{y}$ as
\vspace{-2mm}
\begin{equation*}
\mu _{n}=\int_{-\infty }^{\infty }(y-\bar{y})^{n}\,f(y)\,\mathrm{d}y.
\end{equation*}
\vspace{-3mm}

The moments about its mean are called ``central moments'', and they describe the shape of the function $f(y)$.
The first moment ($\mu_1$) is the mean.
The second central moment ($\mu_2=\bar\Omega^2$) is the variance.\footnote{The average square fluctuation in the energy loss, $\bar\Omega^2=\langle(\delta\varepsilon-\langle\delta\varepsilon \rangle)^2\rangle$, where $\delta\varepsilon$ denotes the energy loss, is the so-called ``straggling parameter''.}
Its positive square root is the standard deviation, and it describes the width $\bar\Omega$
of the distribution. The third ($\mu_3$) and fourth ($\mu_4$) central moments are used to determine the normalized central moments, which define the parameters of asymmetry and
peakedness of the distribution, respectively.

Applied to our problem,
the central moments of energy-loss distribution can be represented in the Mott approximation as follows:
\vspace{-3mm}
\begin{equation}
\label{mun}
\mu_{n,M}=2\pi n_0Z^\prime\delta x\int\limits_{0}^{\pi}[\delta\varepsilon(\vartheta)]^{n}
\frac{d\sigma_\textrm{M}}{d\Omega}\sin\vartheta d\vartheta\,.
\end{equation}
\vspace{-4mm}

\noindent Here, $\delta x$ is the thickness of the layer of matter traversed by the ion.

Using the expression

\vspace{-4mm}

\begin{eqnarray}
(2l+1)xP_{l}^{m}(x)&=&(l+1-m)P_{l+1}^{m}(x)\nonumber\\
&&-(l+m)P_{l-1}^{m}(x)
\end{eqnarray}
\cite{Ryzh} and the orthogonality relation
\vspace{-5mm}

\begin{equation}
\int\limits_{-1}^{1}P_{l_1}^{m}(x)P_{l_2}^{n}(x)dx=\frac{2}{2l_1+1}
\frac{(l_1+m)!}{(l_1-m)!}\delta_{l_1l_2}\,,
\end{equation}

\vspace{-3mm}
\noindent it is not difficult to obtain from (\ref{sigma1}) and (\ref{mun})
the following representations for the quantities
$\mu_{n,M}\; (n=2,3,4)$ in the form of
rapidly converging series:
\begin{eqnarray}
\label{mu2}
\mu_{2,M}&=&a\sum\limits_{l=0}^{\infty}(l+1)\Bigl[\xi^2
\Bigl\vert\frac{F_l}{2l+1}-\frac{F_{l-1}}{2l+3}\Bigl\vert^2\nonumber\\
&&+\Bigl\vert\frac{lF_l}{2l+1}-\frac{(l+2)F_{l+1}}{2l+3}\Bigl\vert^2\Bigl]\,,\\
\label{mu3}
\mu_{3,M}&=&\frac{ap^2}{m}\sum\limits_{l=0}^{\infty}\frac{1}{(2l+1)}
\Bigl[\xi^2\big\vert\tilde F_l\big\vert^2 +l(l+1)\big\vert\tilde G_l\big\vert^2\Bigl]\,,\nonumber
\end{eqnarray}
\vspace{-7mm}
\begin{eqnarray}
\tilde F_l&=&F_l-\frac{l}{2l-1}F_{l-1}-\frac{l+1}{2l+3}F_{l+1},\nonumber\\
\tilde G_l&=&F_l-\frac{l-1}{2l-1}F_{l-1}-\frac{l+2}{2l+3}F_{l+1},
\end{eqnarray}
\begin{eqnarray}
\label{mu4}
\mu_{4,M}&=&a\Bigl(\frac{p^2}{m}\Bigl)^2\sum\limits_{l=0}^{\infty}(l+1)
\Bigl[\xi^2\Bigl\vert\frac{\tilde F_l}{2l+1}-\frac{\tilde F_{l+1}}{2l+3}\Bigl\vert^2\nonumber\\
&&+\Bigl\vert\frac{l\tilde G_l}{2l+1}-\frac{(l+2)\tilde G_{l+1}}{2l+3}\Bigl\vert^2\Bigl],\nonumber
\end{eqnarray}
\begin{equation}
a=2\pi n_0Z^\prime\Delta x(\hbar c\beta)^2/(1-\beta^2).
\end{equation}
It is easily to verify that the terms of these series decrease
as $l^{-2n+1}~(n=\overline{2,4})$ when $l\to \infty$.

In order to assess the applicability of the computation results in
the Born approximation to describing the heavy ion close collisions with the atoms of matter that reads
\begin{eqnarray}
\frac{d\sigma_\textrm{B}}{d\Omega}&=&\nu^2\left(\frac{\hbar}{2p}\right)^2
\frac{1-\beta^2\sin^2(\vartheta/2)}{\sin^4(\vartheta/2)},\\
\label{Born}
\mu_{n,B}&=&\pi n_0Z^\prime\delta x \,\nu^2\Bigl(\frac{\hbar}{m}\Bigl)^2\nonumber\\
&&\times\Bigl(\frac{2p^2}{m}\Bigl)^n\Bigl(\frac{1}{n-1}-\frac{\beta^2}{n}\Bigl)\,,
\end{eqnarray}
the latter
can be compared with the results  of computation  using the formulae (\ref{mu2}), (\ref{mu3}), and (\ref{mu4}).

For this purpose, we introduce the quantities
\begin{equation}
\label{delta}
\delta_n\equiv\delta_n(\mu)=\frac{\mu_{n,M}-\mu_{n,B}}{\mu_{n,B}},\quad n=\overline{2,4},
\end{equation}
which characterize the accuracy of the Born approximation in
computing the higher moments of the heavy ion energy-loss distribution.

Figure 1 and Table 1 show the computation results of (\ref{delta}).
They demonstrate the dependence of $\delta_n(Z,\beta)$  on the ion charge numbers.\\
\begin{figure}[h!]

\begin{center}

\includegraphics[width=1.0\linewidth]{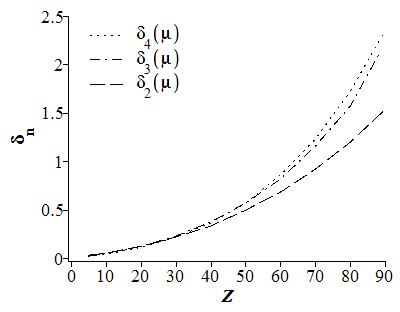}

\caption{Relative corrections  $\delta_n$ to the Born central moments
$\mu_{n,B}$ $(n=\overline{2,4})$ of the energy-loss distribution of incident charged particles as a function of projectile
charge number $Z$ for $\beta = 0.75$.
} \label{Fig1}

\end{center}

\end{figure}

\noindent It is seen from them that the values of $\delta_n(\mu)$ can reach
about two hundred percent for $Z\sim 90$.

Let us notice that the deviation from a simple Born approximation in  determining the ELS
can be very significant.
For instance, the  observed in \cite{2007} energy straggling is about a factor
of 5 lower than the first-order Born result.
The ELS computed in our paper
is generally larger
than that in the Born approximation by a factor of $\sim 2-3$ for heavy ions.
It is consistent with the results of the ELS measurements for relativistic
heavy ions from \cite{GSI2}. The results obtained in our paper
indicate the importance of the MCs
in determining the central moments of the heavy ion energy-loss distributions
at moderate relativistic energies.

\begin{center}
\noindent {\bf Table 1.}
The relative corrections  $\delta_n(\mu)$
to the first-order Born central moments $\mu_{n,B}$ ($n=\overline{2,4}$):\\
$Z$ dependence of $\delta_n(\mu)$ for $\beta = 0.75$.

\bigskip

\begin{tabular}{|c|c|c|c|}
\hline
Z& $\delta_2(\mu)$& $\delta_3(\mu)$& $\delta_4(\mu)$\\
\hline
10.0000& 0.0484& 0.0468& 0.0424\\
20.0000& 0.1172& 0.1185& 0.1112\\
30.0000& 0.2106& 0.2221& 0.2145\\
40.0000& 0.3330& 0.3661& 0.3629\\
50.0000& 0.4891& 0.5607& 0.5696\\
60.0000& 0.6832& 0.8171& 0.8508\\
70.0000& 0.9192& 1.1505& 1.2263\\
80.0000& 1.1988& 1.5727& 1.7190\\
90.0000& 1.5186& 2.0942& 2.3507\\
\hline
\end{tabular}

\end{center}

\bigskip

Thus, the use of the Born approximation in the analysis of close collisions of heavy ions with the atoms of an irradiated material  is at best possible only for the purpose
of qualitative evaluation of the quantities under  consideration.

\section{Calculation of the normalized k-th central moments}
Define now the normalized k-th central moment $\rho_k$ as the k-th central moment divided by
$\bar\Omega^k\equiv\mu_2^{k/2}$.
In doing so, the following relations for the most important normalized 3rd and 4th
central moments of the energy-loss distribution are valid:
\begin{eqnarray}
\rho_{3,M}=\frac{\mu_{3,M}}{\big(\mu_{2,M})^{3/2}}\,,\quad
\rho_{4,M}=\frac{\mu_{4,M}}{\big(\mu_{2,M})^{4/2}}\,.
\end{eqnarray}
The normalized third central moment, the skewness, describes the
stretching of the distribution to the right ($+$) or the left ($-$).
The fourth central moment, the kurtosis, describes the
peakedness of the distribution ($+$, greater than normal, and
$-$, less than normal).

The relative Mott corrections to the normalized first-order Born central moments $\rho_{k,B}$ ($k=3,4$) can be expressed as
\begin{equation}
\delta_k\equiv\delta_k(\rho)=\frac{\rho_{k,M}-\rho_{k,B}}{\rho_{k,B}},\quad k=\overline{3,4}.
\end{equation}

 The calculation results
 for the values of $\delta_k(\rho)$ ($k=\overline{3,4}$)
 are given in Figure 2 and Table 2. They demonstrate the dependence of $\delta_k(\rho)$ on the ion charge number $Z$. These results show that the absolute values of the relative corrections $\delta_3$ to the normalized Born central moment $\rho_{3,B}$ of the relativistic ion energy-loss rate  do reach approximately $20\%$, and the $\vert\delta_4\vert$ values increase to
 $50\%$ for the heavy ions with $Z\sim 90$.

\begin{figure}[h!]
\begin{center}
\includegraphics[width=1.0\linewidth]{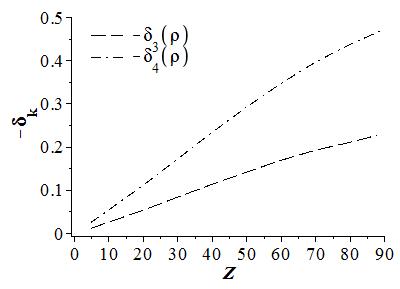}
\caption{\small Relative corrections  $\delta_k$ to the normalized Born central moments
$\rho_{k,B}$ ($k=\overline{3,4}$) of the energy-loss distribution of incident charged particles as a
function of the particle charge number $Z$ for $\beta = 0.75$.
} \label{Fig3}
\end{center}
\end{figure}

\begin{center}
{\bf Table 2.}  The relative corrections $\delta_k(\rho)$
to the normalized first-order Born central moments $\rho_k$:\\
 $Z$ dependence of $\delta_k(\rho)$ for $\beta = 0.75$.

\bigskip

\begin{tabular}{|c|c|c|}
\hline
Z& $-\delta_3(\rho)$& $-\delta_4(\rho)$\\
\hline
10.0000&  \;0.0249& \;0.0516\\
20.0000&  \;0.0529& \;0.1098\\
30.0000&  \;0.0825& \;0.1713\\
40.0000&  \;0.1123& \;0.2330\\
50.0000&  \;0.1411& \;0.2920\\
60.0000&  \;0.1676& \;0.3468\\
70.0000&  \;0.1912& \;0.3956\\
80.0000&  \;0.2109& \;0.4376\\
90.0000&  \;0.2259& \;0.4718\\
\hline
\end{tabular}
\end{center}

\bigskip

The decrease of the Mott normalized central moments ($\rho_{k,M}$)
in comparison with the Born one
($\rho_{k,B}$) for both light and heavy ions means that the ion energy-loss rates,
which were calculated with the use of Mott's corrections, are always less asymmetric
and more ``normal'' (more close to Gaussian) than
those in the Born approximation.

\section{Summary and outlook}

\begin{itemize}

\item
Based on the representation of the Mott corrections $\Delta_\textrm{M} $ to the Bethe-Bloch formula in the form of quite rapidly converging series whose terms are bilinear in the Mott partial amplitudes, an algorithm is proposed to compute the most important central moments $\mu_\textrm {n,M}$ ($n = \overline{2,4} $) of the average energy-loss rates of relativistic charged particles in the Mott approximation.

\item
This algorithm  reduces computing the $\mu_\textrm {n,M}$ to a summing the fast converging (as $ l^{-2n + 1} $ at $ l \to \infty $, $ l\in\mathbb{N}_0$)
infinite series (\ref{mu2}), (\ref{mu3}), (\ref{mu4})  and can be simply implemented using the numerical summation methods of converging series for a given level of precision.

\item

Using the latter result, the parameters of the energy-loss distributions were calculated
for charged particles of moderate relativistic energies
in the Born and Mott approximations for a wide range $ 5 \leq Z \leq 95 $ of the particle charge numbers.

\item The relative Mott corrections $ \delta_\textrm{n} $ and $ \delta_\textrm {k} $ to the Born
values of the central moments $ \mu_\textrm{n,B} $
($n = \overline{2,4}$)
and the normalized central moments $ \rho_\textrm{k,B} $ ($ k = \overline{3,4} $) of the particle average energy-loss distributions in an irradiated material were also computed.

\item It is shown that the relative Mott corrections $ \delta_\textrm{n}(\mu) $ to the Born central moments $ \mu_\textrm{n,B} $ of the distributions can
reach about $200 \% $ for heavy relativistic ions.
This made it possible to explain the results of experiment
\cite{GSI2}.

\item A conclusion is drawn on the applicability of the Born approximation to the consideration of  higher moments of the relativistic heavy ion energy-loss distributions
     at best only for the purpose of their qualitative evaluation.

\item The obtained results also mean that the ion energy-loss distribution, which was computed with the use of Mott's corrections, are always less asymmetric and more normal than the distributions computed in the Born approximation.

\item The choice of an efficient method for numerical calculations of the heavy ion energy-loss distributions and their parameters taking account of the Mott corrections  as well as  comparison of the computational results with the available experimental data will be the subject of a further research.
\end{itemize}

\medskip

\section*{Acknowledgments}

This work was supported by a grant from the Russian Foundation for Basic Research
(project nos. 17-01-00661-а).

}
\end{document}